\documentclass[sn-mathphys,Numbered,iicol]{sn-jnl}% Math and Physical Sciences Reference Style
\usepackage{graphicx}%
\usepackage{multirow}%
\usepackage{amsmath,amssymb,amsfonts}%
\usepackage{amsthm}%
\usepackage{mathrsfs}%
\usepackage[title]{appendix}%
\usepackage{xcolor}%
\usepackage{textcomp}%
\usepackage{manyfoot}%
\usepackage{booktabs}%
\usepackage{algorithm}%
\usepackage{algorithmicx}%
\usepackage{algpseudocode}%
\usepackage{listings}%
\usepackage{orcidlink}
\usepackage{siunitx}
\usepackage{cleveref}% Better referencing of tables and figures etc.
\usepackage{multirow}% For fancy tables
\usepackage{subcaption}% For subtables
\usepackage{array}% Thick table lines
\newcolumntype{?}{!{\vrule width 1.5pt}}

\DeclareSIUnit\angstrom{\text {Å}}
\renewcommand{\citet}[1]{Ref.~\cite{#1}}

\newcommand{\SiGe}{Si$_{1-x}$Ge$_{x}$}
\newcommand{\CB}{C$_{1-x}$B$_{x}$}
\newcommand{\GeSn}{Ge$_{1-x}$Sn$_{x}$}
\newcommand{\GeSi}{Ge$_{1-x}$Si$_{x}$}
\newcommand{\SiC}{Si$_{1-x}$C$_{x}$}

\begin{document}

\title{Dopant concentration effects on \SiGe~crystals for emerging light-source technologies: A molecular dynamics study}

%%=============================================================%%
%% Prefix	-> \pfx{Dr}
%% GivenName	-> \fnm{Joergen W.}
%% Particle	-> \spfx{van der} -> surname prefix
%% FamilyName	-> \sur{Ploeg}
%% Suffix	-> \sfx{IV}
%% NatureName	-> \tanm{Poet Laureate} -> Title after name
%% Degrees	-> \dgr{MSc, PhD}
%% \author*[1,2]{\pfx{Dr} \fnm{Joergen W.} \spfx{van der} \sur{Ploeg} \sfx{IV} \tanm{Poet Laureate}
%%                 \dgr{MSc, PhD}}\email{iauthor@gmail.com}
%%=============================================================%%

\author*{\fnm{Matthew D.} \sur{Dickers\orcidlink{0000-0001-9615-9101}$^{\text{1*}}$}}\email{M.D.Dickers@kent.ac.uk}
%\equalcont{These authors contributed equally to this work.}

\author{\fnm{Gennady B.} \sur{Sushko$^{\text{2}}$}}\email{sushko@mbnexplorer.com}
%\equalcont{These authors contributed equally to this work.}

\author{\fnm{Andrei V.} \sur{Korol\orcidlink{0000-0002-6807-5194}$^{\text{2}}$}}\email{korol@mbnexplorer.com}
%\equalcont{These authors contributed equally to this work.}

\author{\fnm{Nigel J.} \sur{Mason\orcidlink{0000-0002-4468-8324}$^{\text{1}}$}}\email{n.j.mason@kent.ac.uk}
%\equalcont{These authors contributed equally to this work.}

\author{\fnm{Felipe} \sur{Fantuzzi\orcidlink{0000-0002-8200-8262}$^{\text{3}}$}}\email{f.fantuzzi@kent.ac.uk}
%\equalcont{These authors contributed equally to this work.}

\author{\fnm{Andrey V.} \sur{Solov'yov$^{\text{2}}$}}\email{solovyov@mbnresearch.com}
%\equalcont{These authors contributed equally to this work.}

\affil[1]{\orgdiv{School of Physics and Astronomy}, \orgaddress{\street{Ingram Building}, \orgname{University of Kent}, \city{Canterbury}, \postcode{CT2 7NH}, \country{United Kingdom}}}

\affil[2]{\orgname{MBN Research Center}, \orgaddress{\street{Altenh\"oferallee 3}, \postcode{60438} \city{Frankfurt am Main}, \country{Germany}}}

\affil[3]{\orgdiv{School of Chemistry and Forensic Science}, \orgaddress{\street{Ingram Building}, \orgname{University of Kent}, \city{Canterbury}, \postcode{CT2 7NH}, \country{United Kingdom}}}

%%==================================%%
%% sample for unstructured abstract %%
%%==================================%%

\abstract{In this study, we conduct atomistic-level molecular dynamics simulations on fixed-sized silicon-germanium (\SiGe) crystals to elucidate the effects of dopant concentration on the crystalline inter-planar distances. Our calculations consider a range of Ge dopant concentrations between pure Si (0\%) and 15\%, and for both the optimised system state and a temperature of \SI{300}{\kelvin}. We observe a linear relationship between Ge concentration and inter-planar distance and lattice constant, in line with the approximation of Vegard's Law, and other experimental and computational results. These findings will be employed in conjunction with future studies to establish precise tolerances for use in crystal growth, crucial for the manufacture of crystals intended for emerging gamma-ray crystal-based light source technologies.}

\keywords{Doped Crystals, Silicon-Germanium, Inter-planar Distance, Molecular Dynamics}

\maketitle

%%%%%%%%%%%%%%%%%%%%%%%%%%%%%%%%%%%%%%%%%%%%%%%%%%%%%%%%%%%
%%%%%%%%%%%%%%%%%%%%%%%%%%%%%%%%%%%%%%%%%%%%%%%%%%%%%%%%%%%
\section{Introduction} \label{sec:Intro}

Doping has long been used to alter and improve the properties of materials. The introduction of impurities to a base material can lead to significant modification of properties such as electrical conductivity, optical properties, and crystalline structure. In this context, dopant atoms have the capacity to induce structural changes in the material, thus allowing for the precise tuning of the particular properties of a crystal for specific applications. These include existing technologies such as solid-state lasers \cite{Solid_State_Lasers} and high-speed photodetectors \cite{Photodetectors_1, Photodetectors_2, Photodetectors_3}. Doping can also be employed in emerging technologies, including gamma-ray crystal-based light sources (CLSs) \cite{Korol2014Book, Korol2022Book, Korol2020, KOROL2023}.

Gamma-ray CLSs represent a novel, cutting-edge technology designed to generate short-wavelength ($\lambda \ll \SI{1}{\angstrom}$) electromagnetic radiation with high brilliance \cite{Korol2014Book, Korol2020, Korol2022Book, Brilliance}. A detailed explanation of the underlying principles governing their operation is beyond the scope of this current paper. However, these mechanisms are thoroughly described in Refs. \cite{Korol2014Book, Korol2022Book} and the review article \citet{Korol2020}. The key mechanism is the propagation of beams of ultra-relativistic electrons and positrons through oriented crystals (known as \textit{channelling} \cite{Channelling}), leading to the production of radiation of different types.
% The practical realisation of this technology is the focus of international collaboration within the European H2020 N-LIGHT \cite{N-LIGHT} and Horizon European Innovation Council Pathfinder TECHNO-CLS \cite{TECHNO-CLS} projects.

The design of gamma-ray CLSs places a significant emphasis on the structure and quality of the crystals. Low quality crystal structures with many defects lead to dechannelling \cite{Channelling}, in which particles are removed from a channel. Thus, crystals of lower quality will have a shorter dechannelling length: the distance a channelled particle travels prior to dechannelling. Increasing the dechannelling length leads to a subsequent increase in the intensity of the emitted radiation.

\begin{figure}[t!]
    \centering
    \includegraphics[width=0.48\textwidth]{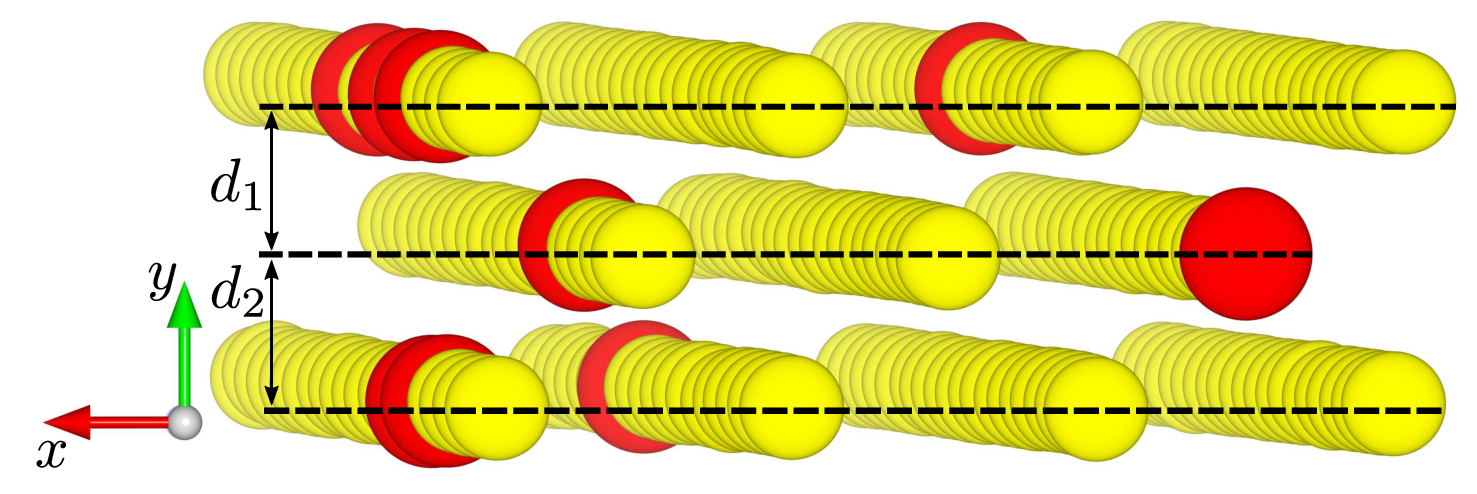}
    \caption{Diagram of a portion of a \SiGe~crystal oriented to highlight the crystalline planes. The inter-planar distances $d_1$ and $d_2$ are shown, and Si atoms are shown in yellow, and Ge atoms in red. Image generated using VESTA \cite{VESTA}.}
    \label{fig:Si-GeFig}
\end{figure}

\begin{figure}[t]
    \centering
    \includegraphics[width=0.47\textwidth]{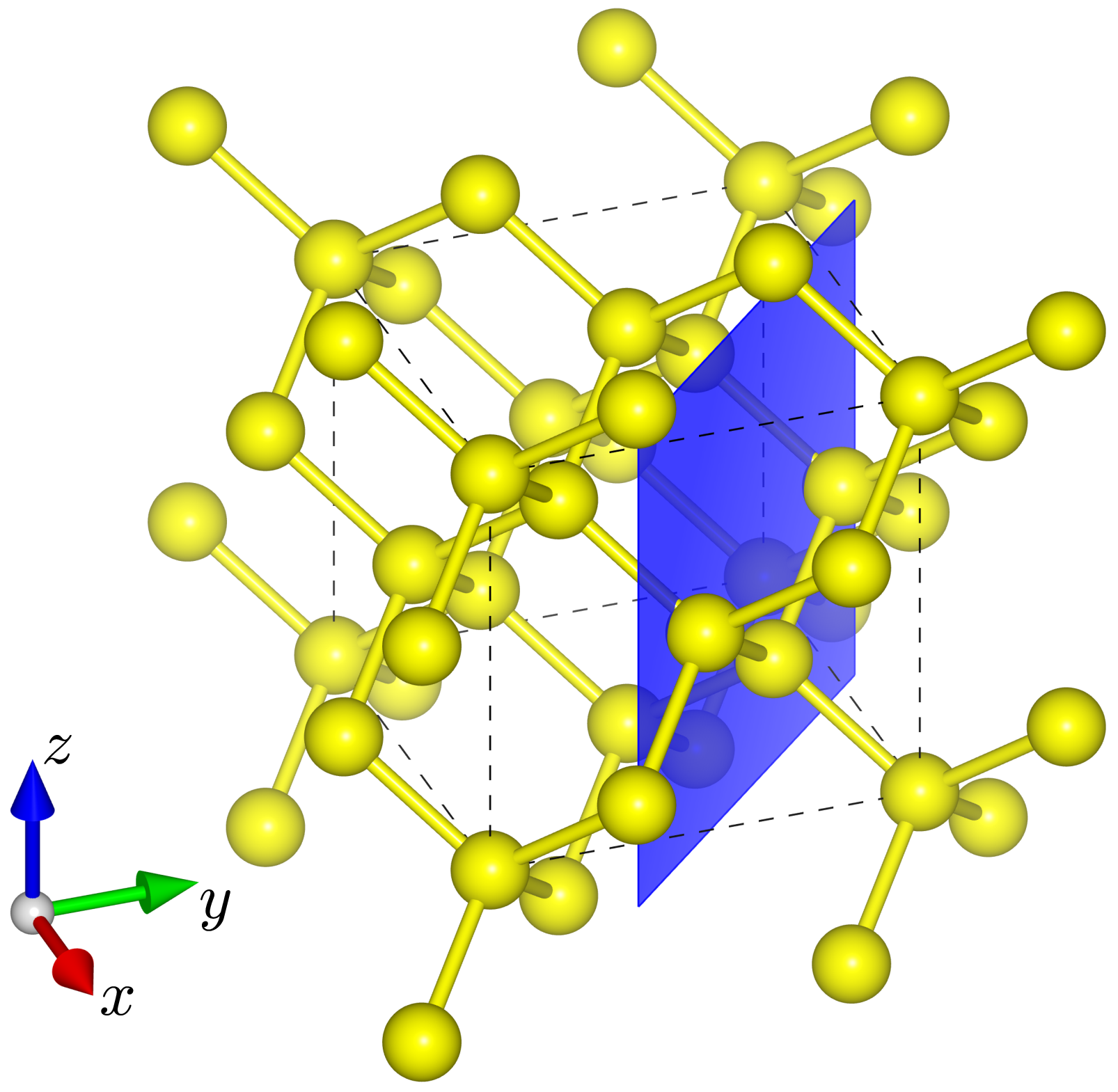}
    \caption{Diagram of a Silicon crystal showing the positions of Si atoms (yellow), the unit cell (dashed lines), and the $(1\;1\;0)$ crystalline plane (blue). Image generated using VESTA \cite{VESTA}.}
    \label{fig:Si_UnitCellFig}
\end{figure}

The crystals used for gamma-ray CLSs can take various forms, including linear, bent, or periodically bent configurations (see Figure 1 in \citet{Korol2020}). Notably, bent and periodically bent crystals offer the unique capability to fine-tune the wavelength and brilliance of the emitted light by adjusting the bending angle. This is attributed to the respective generation of synchrotron-like and undulator-like radiation \cite{InitialCUs, Korol2014Book}, along with the occurrence of channelling radiation \cite{ChannellingRadiation}. The details of these specific mechanisms are described in Refs. \cite{Korol2014Book, Korol2022Book, Korol2020}. Multiple approaches for producing bent and periodically crystals have been developed, including mechanical bending \cite{Compression1, Compression2}, etching \cite{Etching}, scratching\cite{Scratching}, laser ablation \cite{LaserAblation}, acoustic waves \cite{AndreiKorol_1998_acoustic, Acoustic1, Acoustic2, Acoustic3}, and crystal doping \cite{DopedBending}. While all of these techniques can be used for the production of periodically bent crystals, bent crystals are generally produced through surface deformations. Details on the design and production of bent crystals can be found in \citet{Bent_Crystal_Production}. The focus of this study revolves around the controlled periodic bending in crystals through the introduction of dopant atoms.

The principle of producing periodically bent crystals through doping is as follows: A dopant element, with a crystalline lattice constant slightly different from that of the base crystal, is introduced into the base crystal. This mismatch between lattice constants will produce a strain within the crystal, causing a change in the separation $d$ between neighbouring crystalline planes, leading to the bending of crystallographic planes \cite{KRAUSE2002455}. When the dopant element has a larger lattice constant than the base crystal, it leads to an increase of the inter-planar distance. Conversely, a smaller lattice constant causes a decrease in the inter-planar distance. \Cref{fig:Si-GeFig} shows a portion of a Si crystal doped with Ge atoms, and highlights the planes within the crystalline structure and their respective inter-planar distances $d_1$ and $d_2$. These inter-planar distances will depend on the concentration of dopant atoms within each plane. By increasing the concentration of dopant atoms, for example, along the $(1\;0\;0)$ crystal plane, the inter-planar distance is also systematically increased along this plane. Consequently, this will result in crystal bending along the $(1\;1\;0)$ plane. \Cref{fig:Si_UnitCellFig} shows a single Si unit cell, highlighting the $(1\;1\;0)$ plane. Figures 1 and 2 in \citet{BREESE1997540}, and Figure 4 in \citet{KRAUSE2002455} illustrate this bending mechanism. If, at a certain stage, the dopant concentration is intentionally reduced in a systematic manner, the inter-planar distances would likewise decrease, resulting in the formation of a periodically bent structure.

In practice, doped crystals can be fabricated via a number of methods, including diffusion \cite{Diffusion}, ion implantation \cite{IonImplantation}, chemical vapour deposition (CVD) \cite{CVD, Connell2015, BRUNET1998869, WOJEWODA20081302}, and molecular beam epitaxy (MBE) \cite{MBE_SiGe}, with each technique suited to growing different types of crystals. By selecting appropriate atom types and dopant concentrations, periodically bent crystals can be produced. In the case of gamma-ray CLSs, two crystal types have garnered significant research attention: Silicon crystals doped with Germanium (\SiGe), grown through MBE \cite{MBE_SiGe}, and Diamond crystals doped with Boron (\CB), grown via CVD \cite{Connell2015}. Here, $x$ denotes the dopant concentration, where $x=0$ represents the base crystal with no dopant atoms, and $x=1$ denotes a single crystal of the dopant atom.

The dopant concentration can significantly influence the quality of the grown crystal. As dopant atoms are introduced, they induce strain within the crystal, leading to local changes in crystalline structure. These alterations, when considered across the entire crystal, lead to variation in the distances between neighbouring planes. However, if these distances become too large, various types of defects can emerge within the crystalline structure, including point defects and dislocations \cite{DefectBook}. This outcome is dependent on the constituent atoms and the dopant concentration. For example, \CB~crystals have been shown to achieve higher dopant concentrations before defects begin to appear \cite{Diamond1, BRUNET1998869}. In order to successfully manufacture high quality crystals, it is imperative that the maximum dopant concentration is kept below a critical value \cite{KRAUSE2002455}.

Herein, we report the results of an atomisitic-level study of the effect of dopant concentration on defect-free \SiGe~crystals, and specifically how these parameters influence the crystalline properties of inter-planar distance and lattice constant. Our investigation employs molecular dynamics (MD) simulations using the MBN Explorer \cite{MBNExplorer} and MBN Studio \cite{MBNStudio} software packages, allowing for the unique study of the direct effect of dopant atoms on the overall crystalline structure.

This analysis holds significant relevance for the design and practical realisation of gamma-ray CLSs. To ensure the successful production of linear, bent, and periodically bent crystals suitable for use in CLSs, it is essential to identify the manufacturing tolerances of crystal properties. These include the maximum dopant concentration, bending amplitude, bending period, and minimum defect density for which effective channelling is still possible. In this study we focus on the dopant concentration, and the direct impact of dopant atoms on the small-scale crystalline structure.

%%%%%%%%%%%%%%%%%%%%%%%%%%%%%%%%%%%%%%%%%%%%%%%%%%%%%%%%%%%%
\section{Methodology} \label{sec:Method}
This section outlines the computational methodology used to generate defect-free doped \SiGe~crystals, and their subsequent atomistic-level analysis. We considered a fixed crystal size of $(110 \times 110 \times 110)\;\SI{}{\angstrom^3}$, comprising a total of 66,420 atoms. Our simulations were conducted under non-periodic boundary conditions at room temperature, specifically \SI{300}{\kelvin}, consistent with an NVT ensemble. These simulations have been performed using the MBN Explorer software package \cite{MBNExplorer} for advanced multi-scale modelling of complex molecular structures and dynamics. MBN Studio \cite{MBNStudio}, a multi-task toolkit and a dedicated graphical user interface for MBN Explorer, was utilised to construct the systems, prepare input files, and analyse simulation outputs.

Our investigation examines Si crystals doped with Ge at various concentrations to elucidate their influence on the crystalline structure. Due to the lattice mismatch between Si ($a_\text{Si}=\SI{5.431}{\angstrom}$) and Ge  ($a_{\text{Ge}}=\SI{5.658}{\angstrom}$)  \cite{SiGeLatticeConstants}, Ge doping in Si causes an increase in the inter-planar distance. To generate the Si crystals, the MBN Explorer input file was configured with a single Si unit cell. This cell was then systematically duplicated and translated along each axis by the width of one Si lattice constant. This process was repeated until the entire volume of the simulation box was uniformly filled with the crystal structure. Subsequently, a predetermined percentage of Si atoms were selected at random and replaced with Ge atoms, ensuring the amount of Ge atoms aligned with the targeted dopant concentration: between $x=0.00$ (0\%) and $x=0.15$ (15\%). This method initially produces pristine crystalline structures devoid of defects, in comparison to techniques such as MBE and CVD. Although it is feasible to simulate MBE processes through MD calculations \cite{SW_Ge-Ge}, such approaches are computationally expensive and beyond the scope of this study. We abstained from incorporating any form of substrate typical of MBE procedures \cite{MBE_SiGe}, and restricted our examination to a single dopant concentration for each instance. Consequently, the crystals produced in our simulations do not exhibit the distinctive curvature or periodic distortion that are characteristic of periodically bent crystals produced with a gradient in dopant concentration. Subsequent studies will consider these factors in more detail.

The interactions between atoms within the crystals were simulated using the Stillinger$-$Weber potential \cite{StillingerWeberOriginal}. The specific parameters for this potential, along with the atoms involved in these interactions, are summarised in \Cref{tab:StillingerWeberTable}.

\begin{table*}[t!]
    \caption{Tables of the Stillinger$-$Weber potentials used for Si$-$Si and Ge$-$Ge \textbf{(a)}, and Si$-$Ge \textbf{(b)} interactions in Ge doped Si crystals.}
    \begin{subtable}{\textwidth}
        \centering
        \begin{tabular}{c|c|c|c|c|c|c|c|c|c|c}
           Atoms & $A$ & $B$ & $p$ & $q$ & $a$ & $\gamma$ & $\sigma$ (\SI{}{\angstrom}) & $\varepsilon$ (\SI{}{\electronvolt}) & $\lambda$ & Ref. \\ \hline
            Si$-$Si & 7.050 & 0.602 & 4.0 & 0.0 & 1.8 & 1.2 & 2.095 & 2.167 & 21.0 & \cite{StillingerWeberOriginal} \\
            Ge$-$Ge & 7.050 & 0.602 & 4.0 & 0.0 & 1.8 & 1.2 & 2.181 & 1.926 & 31.0 & \cite{SW_Ge-Ge}
        \end{tabular}
        \caption{Table of the Stillinger$-$Weber potential parameters for Si$-$Si and Ge$-$Ge from \citet{StillingerWeberOriginal} and \citet{SW_Ge-Ge} respectively.}
    \end{subtable}

    \begin{subtable}{\textwidth}
        \centering
        \begin{tabular}{c|c?c|c?c|c?c|c}
        \multicolumn{2}{c?}{Constants} & \multicolumn{2}{c?}{$\sigma_{ij}$ (\SI{}{\angstrom})} & \multicolumn{2}{c?}{$\varepsilon_{ijk}$ (\SI{}{\electronvolt})}  & \multicolumn{2}{c}{$\lambda_{ijk}$} \\ \hline
        $A$ & 7.05 & Si$-$Si & 2.095 & Si$-$Si$-$Si & 2.167 & Si$-$Si$-$Si & 21.0 \\
        $B$ & 0.602 & Si$-$Ge & 2.138 & Si$-$Si$-$Ge & 2.104 & Si$-$Si$-$Ge & 23.1 \\
        $p$ & 4.0 & Ge$-$Ge & 2.181 & Si$-$Ge$-$Ge & 2.043 & Si$-$Ge$-$Ge & 25.5 \\
        $q$ & 0.0 &  &  & Ge$-$Si$-$Si & 2.043 & Ge$-$Si$-$Si & 25.5 \\
        $a$ & 1.8 &  &  & Ge$-$Si$-$Ge & 1.984 & Ge$-$Si$-$Ge & 28.1 \\
        $g$ & 1.2 &  &  & Ge$-$Ge$-$Ge & 1.926 & Ge$-$Ge$-$Ge & 31.0
        \end{tabular}
        \caption{Table of the Stillinger$-$Weber potential parameters for Si-Ge from \citet{SW_Ge-Ge}.}
    \end{subtable}
    \label{tab:StillingerWeberTable}
\end{table*}

%%%%%%%%%%%%%%%%%%%%%%%%%%%%%%
For each dopant concentration, $\sim1000$ independent crystalline structures were generated, ensuring a robust and statistically meaningful analysis. These geometries were optimised using the MBN Explorer velocity quenching algorithm over 10000 optimisation steps, and using a simulation box equal to the size of the crystal. Such optimisation simulations represent the crystal structure at some local energy minimum state for a given arrangement of dopant atoms. 50 of these systems were then randomly selected, with the optimised structures as the initial geometries for MD simulations. Each of these systems were heated to \SI{300}{\kelvin} using a Langevin thermostat with damping time of \SI{100}{\femto\second} over a period of \SI{100}{\pico\second}, to allow for thermalisation of the system. To accommodate the expected crystal expansion from atomic rearrangement, the simulation box dimensions were increased to $(120 \times 120 \times 120)\;\SI{}{\angstrom^3}$. At all stages of these simulations, the entire crystal was exposed to vacuum, and no barostat was used.

For our analysis, we used the last 20 frames of the MD simulations, which represent the state of the systems over the concluding \SI{1}{\pico\second}. During this period, we averaged the positions of the atoms in these frames to obtain crystal structures that account for thermal vibrations. To consider the bulk properties of the crystals, we deliberately excluded the edges from our analysis due to their atypical geometry, a consequence of non-periodic boundary interactions in the simulation box. We established a margin of \SI{15}{\angstrom} from each edge of the crystal as the exclusion zone. This specific dimension was chosen as an optimal compromise, minimising the impact of edge artefacts while maximising the core crystal volume available for analysis.

%%%%%%%%%%%%%%%%%%%%%%%%%%%%%%%%%%%%%%%%%%%%%%%%%%%%%%%%%%%%
\section{Results and Discussion} \label{sec:R&D}
As outlined in \Cref{sec:Intro}, the creation of periodically bent \SiGe~crystals is dependent on the change in crystalline inter-planar distances by dopant Ge atoms. By analysing the relationship between the crystalline inter-planar distance and the dopant concentration on an atomistic scale, it is possible to evaluate the range of dopant concentrations for which high-quality, defect-free crystals can be grown. The crystals considered in this study are too small to exhibit the large-scale defect formation that would lead to dechannelling, thus the following analysis is considered from the viewpoint of the direct effect of dopant atoms on the spacing of crystalline planes.

The simulations conducted in this study allow for an atomistic-level investigation into the change in lattice constant and inter-planar distance from the nominal values of \SI{5.431}{\angstrom} and \SI{1.920}{\angstrom}, respectively, for single Si crystal. Due to the cubic structure of the crystal, the lattice constant, hereafter denoted to as $a$, is equal in all lattice families $({\pm1}\;1\;0)$, $(0\;{\pm1}\;0)$, and $(0\;0\;{\pm1})$. Correspondingly, the inter-planar $d(1\;1\;0)$ distance, hereafter denoted as $d$, refers to the separation between adjacent (1\;1\;0) crystalline planes. In order to determine the inter-planar distances within each crystal, atoms that lie along the (1\;1\;0) planes are identified and grouped into their respective planes. A set of coordinates that define the position of each plane relative to all neighbouring planes are then defined by taking the average coordinates in the (1\;1\;0) planes. From this, the distance between each neighbouring plane within the crystal is determined. The inter-planar distance is then averaged over all crystalline planes, and over all simulations for a particular dopant concentration.

\begin{figure*}[t!]
     \centering
     \begin{subfigure}{0.49\textwidth} %0.32 for double column
         \centering
         \includegraphics[width=\textwidth]{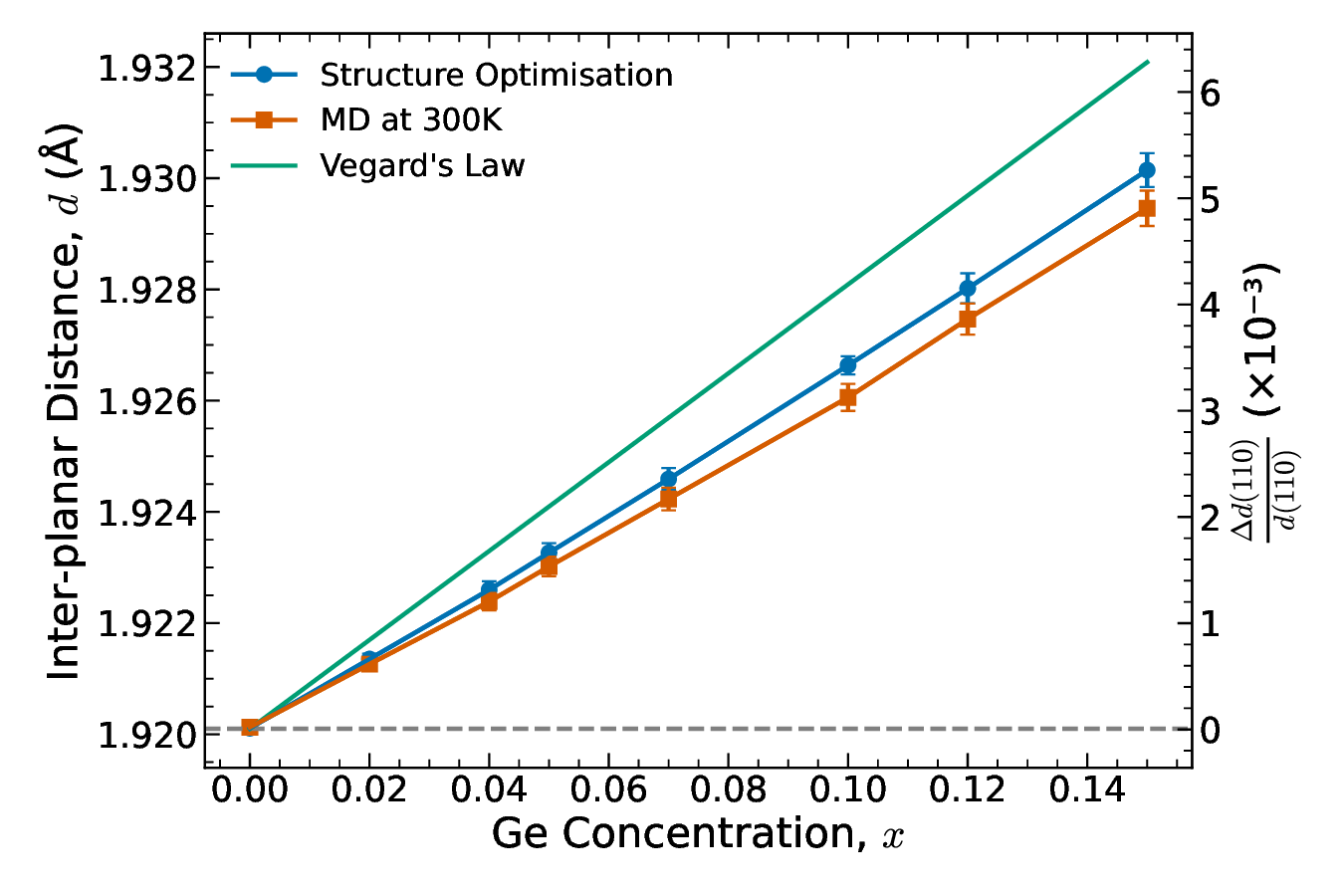}
         \caption{Inter-planar distance}
         \label{subfig:Interplanar_Distance}
     \end{subfigure}
     \hfill
     \begin{subfigure}{0.49\textwidth} %0.32 for double column
         \centering
         \includegraphics[width=\textwidth]{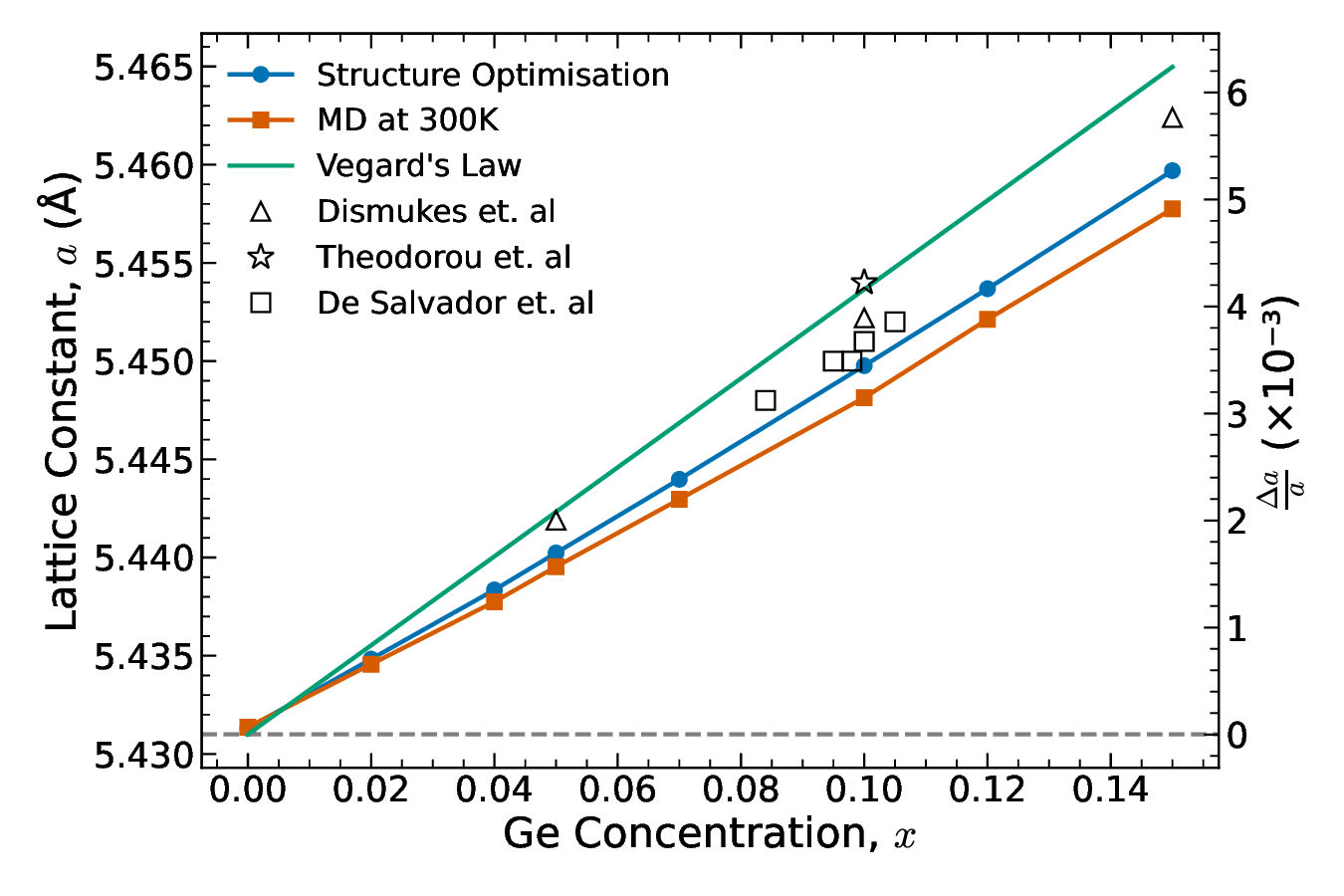}
         \caption{Lattice constant}
         \label{subfig:Lattice_Constant}
     \end{subfigure}
        \caption{Plots of the average inter-planar distance \textbf{(a)} and lattice constant \textbf{(b)} as a function of the dopant concentration. The blue line shows the results of structure optimisation. The orange line shows the results of MD simulations at \SI{300}{\kelvin}. In each plot a second $y$ axis shows the relative change in inter-planar distance $\Delta d(1\;1\;0) / d(1\;1\;0)$ or lattice constant $\Delta a / a$. The lattice constant is compared to experimental data from \citet{Dismukes1964} (open triangles), \citet{Theodorou} (open stars), and \citet{De_Salvador} (open squares), as well as the empirically derived Vegard's Law \cite{Vegard1921} indicated by the green line. The grey dashed lines represent the nominal values of each property in single Si crystals.}
        \label{fig:Plots}
\end{figure*}

Figures \ref{subfig:Interplanar_Distance} and \ref{subfig:Lattice_Constant} show how the average inter-planar distance $d$ and lattice constant $a$ vary with dopant concentration in comparison to the nominal parameters in single Si crystal for both the optimised structure and the system post MD at \SI{300}{\kelvin}. In addition, each plot has a second $y$ axis that shows the relative change in lattice constant $\Delta a / a$ or inter-planar distance $\Delta d(1\;1\;0) / d(1\;1\;0)$, parameters commonly used in the manufacture of crystals \cite{BRUNET1998869, Connell2015}. In both plots we observe an overall linear dependence on inter-planar distance and lattice constant with dopant concentration. These results are compared to those of the empirically derived Vegard's Law \cite{Vegard1921}, predicting a linear proportionality between the lattice constant and the dopant concentration:

\begin{equation}
    a_{\text{SiGe}}(x) = (1-x)a_{\text{Si}} + a_{\text{Ge}}x.
    \label{eqn:Vegards_Law}
\end{equation}

 As the crystal scales in all directions, it follows that the inter-planar distance will adhere to the same relation: $d(x) = (1-x)d_{\text{Si}} + d_{\text{Ge}}x$. Comparing the lines for the optimised structure and that of MD simulations at \SI{300}{\kelvin} it is evident that for increased dopant concentrations the deviation of the lattice constant and inter-planar distances increases. By conducting a linear fit of the form $y=mx+c$, we obtain a relationship comparable to Vegard's Law, \cref{eqn:Vegards_Law}. In the case of \Cref{subfig:Interplanar_Distance}, $c\equiv d_{\text{Si}}=\SI{1.920}{\angstrom}$, and \Cref{subfig:Lattice_Constant}, $c\equiv a_{\text{Si}}=\SI{5.431}{\angstrom}$, equivalent to the nominal parameters for single Si crystals. The values of the gradient $m$ calculated for the inter-planar distance $m_d$ and the lattice constant $m_a$ are as follows:

\begin{gather}
    m_d =
    \begin{cases}
          0.067 & \text{for optimisation} \\
          0.062 & \text{for\;} T=\SI{300}{\kelvin}
    \end{cases}  \ , \\
        m_a =
    \begin{cases}
          0.190 & \text{for optimisation} \\
          0.176 & \text{for\;} T=\SI{300}{\kelvin}
    \end{cases} \ .
\end{gather}

Multiplying the values of $m_d$ by the ratio $a_{\text{Si}} / a_{\text{Ge}}$ yields the values of $m_a$, as expected. This decrease in gradient from optimisation to MD is attributed to the thermalisation of the system, however contradicts what one would expect from Vegard's Law, with the lattice constant of Si corresponding to $a_{\text{Si}}=\SI{5.431}{\angstrom}$ at \SI{300}{\kelvin}. However, such a discrepancy is to be expected; it is well documented that Vegard's Law (which was initially empirically derived) exhibits deviations from the results of both experimental and computational studies of the structure of semiconductor materials. Such deviations have been observed in diffraction experiments of \GeSi~\cite{Dismukes1964, SnGe_Discrepency_Exp} and \GeSn~\cite{SnGe_Discrepency_Exp}, Monte Carlo simulations of \SiGe~\cite{Theodorou} and \GeSi~\cite{Vegard_Disprep_1}, MD simulations of \SiGe~and \SiC~\cite{TersoffVegard}, and DFT simulations of \GeSn~\cite{SnGe_Discrepency_Exp_Theo}. \citet{Vegard_Approximation} analyses Vegard's Law in the frame of thermodynamics, stating that Vegard's Law should be reclassified as an approximation. In particular, they specify that Vegard's Law represents a valid approximation when the lattice constants of the components differ by less than 5\%; in the case of Si and Ge, the lattice constants differ by $\sim4$\%, thus Vegard's Law provides a valid approximation for the change in lattice constant.

Discrepancies from Vegard's Law are often accounted for by the modification of the linear interpolation to a parabolic relation

\begin{equation}
    a_{\text{SiGe}}(x) = (1-x)a_{\text{Si}} + a_{\text{Ge}}x - x(1-x)b_{\text{SiGe}},
\end{equation}

where the first two terms are the the linear relation \cref{eqn:Vegards_Law}, and the final term accounts for the deviation from the linear behaviour, quantified by the bowing parameter $b_{\text{SiGe}}$ \cite{Bowing_Parameter_book}. The values of $b_{\text{SiGe}}$ have been obtained from a parabolic fit:

\begin{equation}
    b_{\text{SiGe}} =
    \begin{cases}
          \SI{0.042}{\angstrom} & \text{for optimisation} \\
          \SI{0.058}{\angstrom} & \text{for\;} T=\SI{300}{\kelvin}
    \end{cases}  \ . \\
\end{equation}

These values are larger compared to those obtained from other studies; $b_{\text{SiGe}} = \SI{0.026}{\angstrom}$ from \citet{Dismukes1964}, $b_{\text{SiGe}} = \SI{0.010}{\angstrom}$ from \citet{Theodorou}, and $b_{\text{SiGe}} = \SI{0.0253}{\angstrom}$ from \citet{SnGe_Discrepency_Exp}. Indeed, the deviation from Vegard's Law we observe is larger than that observed in other studies, seen in \Cref{subfig:Lattice_Constant}. The physical explanation of this discrepancy likely originates in how the crystals used in this study have been generated. In our simulations, the crystals start with an ideal crystalline structure, regardless of the concentration of dopant atoms. The structure of the crystals are then optimised to a local energy minimum state, and are then allowed to relax as MD simulations are conducted. In both types of simulations the crystalline structure rearranges based on the dopant concentration. This is in contrast to crystals grown through MBE or CVD, which allows for the formation of dislocations and defects as the crystal is grown. Defects typically form over size ranges of nanometers \cite{DefectBook}, so it is not unexpected that our $\sim\SI{100}{\angstrom}$ crystals show an ideal structure. In addition, our simulations neglect the substrates typically used for growth of crystals via MBE and CVD, and in combination with the non-peridoic nature of the simulation box, the crystals are allowed to scale in all directions, rather then just perpendicular to the substrate. This will impact the overall lattice constant.

%%%%%%%%%%%%%%%%%%%%%%%%%%%%%%%%%%%%%%%%%%%%%%%%%%%%%%%%%%%%
 \section{Conclusions and Outlook} \label{sec:Conclusions}

In this study, we have conducted atomistic-level simulations to examine the influence of dopant atoms on the structure of small \SiGe~crystals, all within the context of emerging technologies for gamma-ray crystal-based light sources (CLSs). We have shown a linear relationship between dopant concentration and inter-planar distance and lattice constant between $x=0.00$ and $x=0.15$.
The thermalisation of the system affects the gradient of the linear relationship, with MD simulations conducted at \SI{300}{\kelvin} resulting in a reduced lattice constant as compared to the optimised structure. Our results show similar deviations from Vegard's Law to Refs.~\cite{Dismukes1964, SnGe_Discrepency_Exp, Theodorou, Vegard_Disprep_1, TersoffVegard, SnGe_Discrepency_Exp_Theo}, but are in agreement with Vegard's Law when it is considered as a general approximation, as outlined in \citet{Vegard_Approximation}. We observe an overall smaller lattice constant than \citet{Dismukes1964} and \cite{Theodorou}; however, we attribute this to our crystal generation method. Our crystals, being smaller and initially possessing an ideal structure, exhibit a much closer resemblance to an ideal crystal following MD simulations in comparison to the studies referenced. Their small size means large-scale defect formation is not possible, unlike crystals grown through conventional means. In addition, expansion of the crystal is not limited to a particular direction, as would be the case if a substrate were present.

This work has developed the basis for the simulation protocol for generating doped crystals in the context of gamma-ray CLSs. Future studies considering additional parameters, crystal types, and sizes will allow for the investigation of defect formation, as well as the influence of the substrate. Furthermore, the methodology outlined in this work allows for the analysis of various aspects such as crystal amorphisation, thermal expansion, thermal conductivity, and more.

The quality of crystals and understanding of defect forming processes are essential for the creation of gamma-ray CLSs. These simulations may be coupled with experimental studies to characterise the quality of candidate crystals, and channelling experiments run in parallel with relativistic MD \cite{RelativisticMD} channelling simulations to evaluate the efficacy of radiation production of particular crystals. In the future more accurate crystal growth methods can be considered, such as MBE and CVD. These can be effectively explored using stochastic processes, such as Kinetic Monte Carlo simulations, which model the crystal growth through probabilistic processes \cite{VOIGTLANDER2001127}. This approach enables simulations over significantly longer timescales than traditional MD, making it comparable to MBE and CVD crystal growth. The recent successful integration of stochastic dynamics into MBN Explorer \cite{Stochastic_MBN} has opened the door to further investigations into such stochastic processes.

This research field is inherently multidisciplinary, integrating simulation and experimental studies that are essential for the realisation of emerging gamma-ray CLSs.

%%%%%%%%%%%%%%%%%%%%%%%%%%%%%%%%%%%%%%%%%%%%%%%%%%%%%%%%%%%
%%%%%%%%%%%%%%%%%%%%%%%%%%%%%%%%%%%%%%%%%%%%%%%%%%%%%%%%%%%
\section*{Statements and declarations}

\bmhead{Acknowledgments}
The authors acknowledge financial support from the European Commission's Horizon Europe-EIC-Pathfinder-Open TECHNO-CLS (G.A. 101046458) project, from the H2020 RISE-NLIGHT project (G.A. 872196), and from the COST Action
CA20129 MultIChem, supported by COST (European Cooperation in Science and Technology).
The work was supported in part by Deutsche Forschungsgemeinschaft, Germany (Project No. 413220201).
M. D. Dickers gratefully acknowledges receipt of a postgraduate studentship from UKRI.
The possibility of performing computer simulations at the Goethe-HLR cluster of the Frankfurt Center for Scientific Computing is gratefully acknowledged.
The authors kindly thank A. V. Verkhovtsev for fruitful discussions and the anonymous referees, whose valuable comments have improved the quality and clarity of this manuscript.

\bmhead{Competing Interests}
The authors do not declare any conflicts of interest, and there is no financial interest to report.

\bmhead{Author Contribution}
MDD and AVK conducted the computational simulations and data analysis. MDD wrote the manuscript with support from AVK and input from all authors. GBS provided technical and computational advice and support for these simulations. NJM, FF, and AVS helped supervise the project and all authors provided critical feedback and helped shape the research and analysis.

\bmhead{Data Availability}
The datasets generated and/or analysed during the current study are available from the corresponding author upon reasonable request.

\bibliography{sn-bibliography.bib}

%% BioMed_Central_Bib_Style_v1.01

\begin{thebibliography}{53}
% BibTex style file: bmc-mathphys.bst (version 2.1), 2014-07-24
\ifx \bisbn   \undefined \def \bisbn  #1{ISBN #1}\fi
\ifx \binits  \undefined \def \binits#1{#1}\fi
\ifx \bauthor  \undefined \def \bauthor#1{#1}\fi
\ifx \batitle  \undefined \def \batitle#1{#1}\fi
\ifx \bjtitle  \undefined \def \bjtitle#1{#1}\fi
\ifx \bvolume  \undefined \def \bvolume#1{\textbf{#1}}\fi
\ifx \byear  \undefined \def \byear#1{#1}\fi
\ifx \bissue  \undefined \def \bissue#1{#1}\fi
\ifx \bfpage  \undefined \def \bfpage#1{#1}\fi
\ifx \blpage  \undefined \def \blpage #1{#1}\fi
\ifx \burl  \undefined \def \burl#1{\textsf{#1}}\fi
\ifx \doiurl  \undefined \def \doiurl#1{\url{https://doi.org/#1}}\fi
\ifx \betal  \undefined \def \betal{\textit{et al.}}\fi
\ifx \binstitute  \undefined \def \binstitute#1{#1}\fi
\ifx \binstitutionaled  \undefined \def \binstitutionaled#1{#1}\fi
\ifx \bctitle  \undefined \def \bctitle#1{#1}\fi
\ifx \beditor  \undefined \def \beditor#1{#1}\fi
\ifx \bpublisher  \undefined \def \bpublisher#1{#1}\fi
\ifx \bbtitle  \undefined \def \bbtitle#1{#1}\fi
\ifx \bedition  \undefined \def \bedition#1{#1}\fi
\ifx \bseriesno  \undefined \def \bseriesno#1{#1}\fi
\ifx \blocation  \undefined \def \blocation#1{#1}\fi
\ifx \bsertitle  \undefined \def \bsertitle#1{#1}\fi
\ifx \bsnm \undefined \def \bsnm#1{#1}\fi
\ifx \bsuffix \undefined \def \bsuffix#1{#1}\fi
\ifx \bparticle \undefined \def \bparticle#1{#1}\fi
\ifx \barticle \undefined \def \barticle#1{#1}\fi
\bibcommenthead
\ifx \bconfdate \undefined \def \bconfdate #1{#1}\fi
\ifx \botherref \undefined \def \botherref #1{#1}\fi
\ifx \url \undefined \def \url#1{\textsf{#1}}\fi
\ifx \bchapter \undefined \def \bchapter#1{#1}\fi
\ifx \bbook \undefined \def \bbook#1{#1}\fi
\ifx \bcomment \undefined \def \bcomment#1{#1}\fi
\ifx \oauthor \undefined \def \oauthor#1{#1}\fi
\ifx \citeauthoryear \undefined \def \citeauthoryear#1{#1}\fi
\ifx \endbibitem  \undefined \def \endbibitem {}\fi
\ifx \bconflocation  \undefined \def \bconflocation#1{#1}\fi
\ifx \arxivurl  \undefined \def \arxivurl#1{\textsf{#1}}\fi
\csname PreBibitemsHook\endcsname

%%% 1
\bibitem[\protect\citeauthoryear{Denker and
  Shklovsky}{2013}]{Solid_State_Lasers}
\begin{bbook}
\bauthor{\bsnm{Denker}, \binits{B.}},
\bauthor{\bsnm{Shklovsky}, \binits{E.}}:
\bbtitle{{Handbook of Solid-state Lasers: Materials, Systems and
  Applications}}.
\bpublisher{Woodhead Publishing},
\blocation{Cham}
(\byear{2013}).
\doiurl{10.1533/9780857097507.2.171}
\end{bbook}
\endbibitem

%%% 2
\bibitem[\protect\citeauthoryear{Ottaviano et~al.}{2012}]{Photodetectors_1}
\begin{bchapter}
\bauthor{\bsnm{Ottaviano}, \binits{L.}},
\bauthor{\bsnm{Semenova}, \binits{E.}},
\bauthor{\bsnm{Schubert}, \binits{M.}},
\bauthor{\bsnm{Yvind}, \binits{K.}},
\bauthor{\bsnm{Armaroli}, \binits{A.}},
\bauthor{\bsnm{Bellanca}, \binits{G.}},
\bauthor{\bsnm{Trillo}, \binits{S.}},
\bauthor{\bsnm{Nguyen}, \binits{T.N.}},
\bauthor{\bsnm{Gay}, \binits{M.}},
\bauthor{\bsnm{Bramerie}, \binits{L.}},
\bauthor{\bsnm{Simon}, \binits{J.-C.}}:
\bctitle{{High-speed photodetectors in a photonic crystal platform}}.
In: \bbtitle{{2012 Conference on Lasers and Electro-Optics (CLEO)}},
pp. \bfpage{1}--\blpage{2}
(\byear{2012}).
\doiurl{10.1364/CLEO_SI.2012.CM1A.2}
\end{bchapter}
\endbibitem

%%% 3
\bibitem[\protect\citeauthoryear{Shkir et~al.}{2019}]{Photodetectors_2}
\begin{barticle}
\bauthor{\bsnm{Shkir}, \binits{M.}},
\bauthor{\bsnm{Khan}, \binits{M.T.}},
\bauthor{\bsnm{Ashraf}, \binits{I.M.}},
\bauthor{\bsnm{Almohammedi}, \binits{A.}},
\bauthor{\bsnm{Dieguez}, \binits{E.}},
\bauthor{\bsnm{AlFaify}, \binits{S.}}:
\batitle{{High-performance visible light photodetectors based on inorganic CZT
  and InCZT single crystals}}.
\bjtitle{Sci. Rep.}
\bvolume{9}(\bissue{1}),
\bfpage{12436}
(\byear{2019})
\doiurl{10.1038/s41598-019-48621-3}
\end{barticle}
\endbibitem

%%% 4
\bibitem[\protect\citeauthoryear{Jethwa et~al.}{2021}]{Photodetectors_3}
\begin{barticle}
\bauthor{\bsnm{Jethwa}, \binits{V.P.}},
\bauthor{\bsnm{Patel}, \binits{K.}},
\bauthor{\bsnm{Pathak}, \binits{V.M.}},
\bauthor{\bsnm{Solanki}, \binits{G.K.}}:
\batitle{{Enhanced electrical and optoelectronic performance of SnS crystal by
  Se doping}}.
\bjtitle{J. Alloys Compd.}
\bvolume{883},
\bfpage{160941}
(\byear{2021})
\doiurl{10.1016/j.jallcom.2021.160941}
\end{barticle}
\endbibitem

%%% 5
\bibitem[\protect\citeauthoryear{Korol et~al.}{2014}]{Korol2014Book}
\begin{bbook}
\bauthor{\bsnm{Korol}, \binits{A.V.}},
\bauthor{\bsnm{Solov'yov}, \binits{A.V.}},
\bauthor{\bsnm{Greiner}, \binits{W.}}:
\bbtitle{{Channeling and Radiation in Periodically Bent Crystals}}.
\bpublisher{Springer},
\blocation{Heidelberg}
(\byear{2014}).
\doiurl{10.1007/978-3-642-54933-5}
\end{bbook}
\endbibitem

%%% 6
\bibitem[\protect\citeauthoryear{Korol and Solov'yov}{2022}]{Korol2022Book}
\begin{bbook}
\bauthor{\bsnm{Korol}, \binits{A.}},
\bauthor{\bsnm{Solov'yov}, \binits{A.V.}}:
\bbtitle{{Novel Lights Sources Beyond Free Electron Lasers}}.
\bpublisher{Springer},
\blocation{Cham}
(\byear{2022}).
\doiurl{10.1007/978-3-031-04282-9}
\end{bbook}
\endbibitem

%%% 7
\bibitem[\protect\citeauthoryear{Korol and Solov'yov}{2020}]{Korol2020}
\begin{barticle}
\bauthor{\bsnm{Korol}, \binits{A.V.}},
\bauthor{\bsnm{Solov'yov}, \binits{A.V.}}:
\batitle{{Crystal-based Intensive Gamma-ray Light Sources}}.
\bjtitle{Europ. Phys. J. D.}
\bvolume{74}(\bissue{10}),
\bfpage{201}
(\byear{2020})
\doiurl{10.1140/epjd/e2020-10239-8}
\end{barticle}
\endbibitem

%%% 8
\bibitem[\protect\citeauthoryear{Korol and Solov’yov}{2023}]{KOROL2023}
\begin{barticle}
\bauthor{\bsnm{Korol}, \binits{A.V.}},
\bauthor{\bsnm{Solov’yov}, \binits{A.V.}}:
\batitle{{Atomistic Modeling and Characterizaion of Light Sources Based on
  Small-amplitude Short-period Periodically Bent Crystals}}.
\bjtitle{Nucl. Instrum. Meth. B}
\bvolume{537},
\bfpage{1}--\blpage{13}
(\byear{2023})
\doiurl{10.1016/j.nimb.2023.01.012}
\end{barticle}
\endbibitem

%%% 9
\bibitem[\protect\citeauthoryear{Sushko et~al.}{2022}]{Brilliance}
\begin{barticle}
\bauthor{\bsnm{Sushko}, \binits{G.B.}},
\bauthor{\bsnm{Korol}, \binits{A.V.}},
\bauthor{\bsnm{Solov'yov}, \binits{A.V.}}:
\batitle{{Extremely Brilliant Crystal-based Light Sources}}.
\bjtitle{Europ. Phys. J. D.}
\bvolume{76}(\bissue{9}),
\bfpage{166}
(\byear{2022})
\doiurl{10.1140/epjd/s10053-022-00502-7}
\end{barticle}
\endbibitem

%%% 10
\bibitem[\protect\citeauthoryear{Lindhard}{1965}]{Channelling}
\begin{barticle}
\bauthor{\bsnm{Lindhard}, \binits{J.}}:
\batitle{{Influence Of Crystal Lattice On Motion Of Energetic Charged
  Particles}}.
\bjtitle{{Kongel. Dan. Vidensk. Selsk., Mat.-Fys. Medd.}}
\bvolume{34}(\bissue{14}),
\bfpage{1}--\blpage{64}
(\byear{1965})
\end{barticle}
\endbibitem

%%% 11
\bibitem[\protect\citeauthoryear{Momma and Izumi}{2011}]{VESTA}
\begin{barticle}
\bauthor{\bsnm{Momma}, \binits{K.}},
\bauthor{\bsnm{Izumi}, \binits{F.}}:
\batitle{{{\it VESTA3} for three-dimensional visualization of crystal,
  volumetric and morphology data}}.
\bjtitle{J. Appl. Crystallogr.}
\bvolume{44}(\bissue{6}),
\bfpage{1272}--\blpage{1276}
(\byear{2011})
\doiurl{10.1107/S0021889811038970}
\end{barticle}
\endbibitem

%%% 12
\bibitem[\protect\citeauthoryear{Korol et~al.}{1999}]{InitialCUs}
\begin{barticle}
\bauthor{\bsnm{Korol}, \binits{A.V.}},
\bauthor{\bsnm{Solov'yov}, \binits{A.V.}},
\bauthor{\bsnm{Greiner}, \binits{W.}}:
\batitle{{Photon Emission by an Ultra-relativistic Particle Channeling in a
  Periodically Bent Crystal}}.
\bjtitle{Int. J. Mod. Phys. E}
\bvolume{08}(\bissue{01}),
\bfpage{49}--\blpage{100}
(\byear{1999})
\doiurl{10.1142/S0218301399000069}
\end{barticle}
\endbibitem

%%% 13
\bibitem[\protect\citeauthoryear{Kumakhov}{1976}]{ChannellingRadiation}
\begin{barticle}
\bauthor{\bsnm{Kumakhov}, \binits{M.A.}}:
\batitle{{On the Theory of Electromagnetic Radiation of Charged Particles in a
  Crystal}}.
\bjtitle{Phys. Lett. A}
\bvolume{57}(\bissue{1}),
\bfpage{17}--\blpage{18}
(\byear{1976})
\doiurl{10.1016/0375-9601(76)90438-2}
\end{barticle}
\endbibitem

%%% 14
\bibitem[\protect\citeauthoryear{Guidi et~al.}{2011}]{Compression1}
\begin{barticle}
\bauthor{\bsnm{Guidi}, \binits{V.}},
\bauthor{\bsnm{Lanzoni}, \binits{L.}},
\bauthor{\bsnm{Mazzolari}, \binits{A.}}:
\batitle{{Patterning and Modeling of Mechanically Bent Silicon Plates Deformed
  through Coactive Stresses}}.
\bjtitle{Thin Solid Films}
\bvolume{520}(\bissue{3}),
\bfpage{1074}--\blpage{1079}
(\byear{2011})
\doiurl{10.1016/j.tsf.2011.09.008}
\end{barticle}
\endbibitem

%%% 15
\bibitem[\protect\citeauthoryear{Guidi et~al.}{2007}]{Compression2}
\begin{barticle}
\bauthor{\bsnm{Guidi}, \binits{V.}},
\bauthor{\bsnm{Lanzoni}, \binits{L.}},
\bauthor{\bsnm{Mazzolari}, \binits{A.}},
\bauthor{\bsnm{Martinelli}, \binits{G.}},
\bauthor{\bsnm{Tralli}, \binits{A.}}:
\batitle{{Design of a Crystalline Undulator Based on Patterning by Tensile
  Si3N4 Strips on a Si Crystal}}.
\bjtitle{Appl. Phys. Lett.}
\bvolume{90}(\bissue{11}),
\bfpage{114107}
(\byear{2007})
\doiurl{10.1063/1.2712510}
\end{barticle}
\endbibitem

%%% 16
\bibitem[\protect\citeauthoryear{Guidi et~al.}{2005}]{Etching}
\begin{barticle}
\bauthor{\bsnm{Guidi}, \binits{V.}},
\bauthor{\bsnm{Antonini}, \binits{A.}},
\bauthor{\bsnm{Baricordi}, \binits{S.}},
\bauthor{\bsnm{Logallo}, \binits{F.}},
\bauthor{\bsnm{Malagù}, \binits{C.}},
\bauthor{\bsnm{Milan}, \binits{E.}},
\bauthor{\bsnm{Ronzoni}, \binits{A.}},
\bauthor{\bsnm{Stefancich}, \binits{M.}},
\bauthor{\bsnm{Martinelli}, \binits{G.}},
\bauthor{\bsnm{Vomiero}, \binits{A.}}:
\batitle{{Tailoring of Silicon Crystals for Relativistic-particle Channeling}}.
\bjtitle{Nucl. Instrum. Meth. B}
\bvolume{234}(\bissue{1}),
\bfpage{40}--\blpage{46}
(\byear{2005})
\doiurl{10.1016/j.nimb.2005.01.008}
\end{barticle}
\endbibitem

%%% 17
\bibitem[\protect\citeauthoryear{Bellucci et~al.}{2003}]{Scratching}
\begin{barticle}
\bauthor{\bsnm{Bellucci}, \binits{S.}},
\bauthor{\bsnm{Bini}, \binits{S.}},
\bauthor{\bsnm{Biryukov}, \binits{V.M.}},
\bauthor{\bsnm{Chesnokov}, \binits{Y.A.}},
\bauthor{\bsnm{Dabagov}, \binits{S.}},
\bauthor{\bsnm{Giannini}, \binits{G.}},
\bauthor{\bsnm{Guidi}, \binits{V.}},
\bauthor{\bsnm{Ivanov}, \binits{Y.M.}},
\bauthor{\bsnm{Kotov}, \binits{V.I.}},
\bauthor{\bsnm{Maisheev}, \binits{V.A.}},
\bauthor{\bsnm{Malag\`u}, \binits{C.}},
\bauthor{\bsnm{Martinelli}, \binits{G.}},
\bauthor{\bsnm{Petrunin}, \binits{A.A.}},
\bauthor{\bsnm{Skorobogatov}, \binits{V.V.}},
\bauthor{\bsnm{Stefancich}, \binits{M.}},
\bauthor{\bsnm{Vincenzi}, \binits{D.}}:
\batitle{{Experimental Study for the Feasibility of a Crystalline Undulator}}.
\bjtitle{Phys. Rev. Lett.}
\bvolume{90}(\bissue{3}),
\bfpage{034801}
(\byear{2003})
\doiurl{10.1103/PhysRevLett.90.034801}
\end{barticle}
\endbibitem

%%% 18
\bibitem[\protect\citeauthoryear{Balling et~al.}{2009}]{LaserAblation}
\begin{barticle}
\bauthor{\bsnm{Balling}, \binits{P.}},
\bauthor{\bsnm{Esberg}, \binits{J.}},
\bauthor{\bsnm{Kirsebom}, \binits{K.}},
\bauthor{\bsnm{Le}, \binits{D.Q.S.}},
\bauthor{\bsnm{Uggerhøj}, \binits{U.I.}},
\bauthor{\bsnm{Connell}, \binits{S.H.}},
\bauthor{\bsnm{Härtwig}, \binits{J.}},
\bauthor{\bsnm{Masiello}, \binits{F.}},
\bauthor{\bsnm{Rommeveaux}, \binits{A.}}:
\batitle{{Bending Diamonds by Femtosecond Laser Ablation}}.
\bjtitle{Nucl. Instrum. Meth. B}
\bvolume{267}(\bissue{17}),
\bfpage{2952}--\blpage{2957}
(\byear{2009})
\doiurl{10.1016/j.nimb.2009.06.109}
\end{barticle}
\endbibitem

%%% 19
\bibitem[\protect\citeauthoryear{Korol
  et~al.}{1998}]{AndreiKorol_1998_acoustic}
\begin{barticle}
\bauthor{\bsnm{Korol}, \binits{A.V.}},
\bauthor{\bsnm{Solov'yov}, \binits{A.V.}},
\bauthor{\bsnm{Greiner}, \binits{W.}}:
\batitle{{Coherent Radiation of an Ultrarelativistic Charged Particle
  Channelled in a Periodically Bent Crystal}}.
\bjtitle{J. Phys. G}
\bvolume{24}(\bissue{5}),
\bfpage{45}
(\byear{1998})
\doiurl{10.1088/0954-3899/24/5/001}
\end{barticle}
\endbibitem

%%% 20
\bibitem[\protect\citeauthoryear{Wagner et~al.}{2011}]{Acoustic1}
\begin{barticle}
\bauthor{\bsnm{Wagner}, \binits{W.}},
\bauthor{\bsnm{Azadegan}, \binits{B.}},
\bauthor{\bsnm{Buettig}, \binits{H.}},
\bauthor{\bsnm{Grigoryan}, \binits{L.S.}},
\bauthor{\bsnm{Mkrtchyan}, \binits{A.}},
\bauthor{\bsnm{Pawelke}, \binits{J.}}:
\batitle{{Channeling Radiation on Quartz Stimulated by Acoustic Waves}}.
\bjtitle{Nuovo Cimento C}
\bvolume{34}(\bissue{4}),
\bfpage{133}--\blpage{140}
(\byear{2011})
\doiurl{10.1393/ncc/i2011-10899-4}
\end{barticle}
\endbibitem

%%% 21
\bibitem[\protect\citeauthoryear{Dedkov}{1994}]{Acoustic2}
\begin{barticle}
\bauthor{\bsnm{Dedkov}, \binits{G.V.}}:
\batitle{{Channeling Radiation in a Crystal Undergoing an Action of Ultrasonic
  or Electromagnetic Waves}}.
\bjtitle{Phys. Stat. Sol.}
\bvolume{184}(\bissue{2}),
\bfpage{535}--\blpage{542}
(\byear{1994})
\doiurl{10.1002/pssb.2221840227}
\end{barticle}
\endbibitem

%%% 22
\bibitem[\protect\citeauthoryear{Ikezi et~al.}{1984}]{Acoustic3}
\begin{barticle}
\bauthor{\bsnm{Ikezi}, \binits{H.}},
\bauthor{\bsnm{Lin-Liu}, \binits{Y.}},
\bauthor{\bsnm{Ohkawa}, \binits{T.}}:
\batitle{{Channeling Radiation in a Periodically Distorted Crystal}}.
\bjtitle{Phys. Rev. B}
\bvolume{30}(\bissue{3}),
\bfpage{1567}--\blpage{1569}
(\byear{1984})
\doiurl{10.1103/PhysRevB.30.1567}
\end{barticle}
\endbibitem

%%% 23
\bibitem[\protect\citeauthoryear{Mikkelsen and Uggerhøj}{2000}]{DopedBending}
\begin{barticle}
\bauthor{\bsnm{Mikkelsen}, \binits{U.}},
\bauthor{\bsnm{Uggerhøj}, \binits{E.}}:
\batitle{{A Crystalline Undulator Based on Graded Composition Strained Layers
  in a Superlattice}}.
\bjtitle{Nucl. Instrum. Meth. B}
\bvolume{160}(\bissue{3}),
\bfpage{435}--\blpage{439}
(\byear{2000})
\doiurl{10.1016/S0168-583X(99)00637-0}
\end{barticle}
\endbibitem

%%% 24
\bibitem[\protect\citeauthoryear{Romagnoni
  et~al.}{2022}]{Bent_Crystal_Production}
\begin{barticle}
\bauthor{\bsnm{Romagnoni}, \binits{M.}},
\bauthor{\bsnm{Guidi}, \binits{V.}},
\bauthor{\bsnm{Bandiera}, \binits{L.}},
\bauthor{\bsnm{De~Salvador}, \binits{D.}},
\bauthor{\bsnm{Mazzolari}, \binits{A.}},
\bauthor{\bsnm{Sgarbossa}, \binits{F.}},
\bauthor{\bsnm{Soldani}, \binits{M.}},
\bauthor{\bsnm{Sytov}, \binits{A.}},
\bauthor{\bsnm{Tamisari}, \binits{M.}}:
\batitle{{Bent Crystal Design and Characterization for High-Energy Physics
  Experiments}}.
\bjtitle{Crystals}
\bvolume{12}(\bissue{9}),
\bfpage{1263}
(\byear{2022})
\doiurl{10.3390/cryst12091263}
\end{barticle}
\endbibitem

%%% 25
\bibitem[\protect\citeauthoryear{Krause et~al.}{2002}]{KRAUSE2002455}
\begin{barticle}
\bauthor{\bsnm{Krause}, \binits{W.}},
\bauthor{\bsnm{Korol}, \binits{A.V.}},
\bauthor{\bsnm{Solov'yov}, \binits{A.V.}},
\bauthor{\bsnm{Greiner}, \binits{W.}}:
\batitle{{Photon Emission by Ultra-relativistic Positrons in Crystalline
  Undulators: The High-energy Regime}}.
\bjtitle{Nucl. Instrum. Meth. A}
\bvolume{483}(\bissue{1}),
\bfpage{455}--\blpage{460}
(\byear{2002})
\doiurl{10.1016/S0168-9002(02)00361-3}
\end{barticle}
\endbibitem

%%% 26
\bibitem[\protect\citeauthoryear{Breese}{1997}]{BREESE1997540}
\begin{barticle}
\bauthor{\bsnm{Breese}, \binits{M.B.H.}}:
\batitle{{Beam Bending using Graded Composition Strained Layers}}.
\bjtitle{Nucl. Instrum. Meth. B}
\bvolume{132}(\bissue{3}),
\bfpage{540}--\blpage{547}
(\byear{1997})
\doiurl{10.1016/S0168-583X(97)00455-2}
\end{barticle}
\endbibitem

%%% 27
\bibitem[\protect\citeauthoryear{Vlaskin et~al.}{2013}]{Diffusion}
\begin{barticle}
\bauthor{\bsnm{Vlaskin}, \binits{V.A.}},
\bauthor{\bsnm{Barrows}, \binits{C.J.}},
\bauthor{\bsnm{Erickson}, \binits{C.S.}},
\bauthor{\bsnm{Gamelin}, \binits{D.R.}}:
\batitle{{Nanocrystal Diffusion Doping}}.
\bjtitle{J. Am. Chem. Soc.}
\bvolume{135}(\bissue{38}),
\bfpage{14380}--\blpage{14389}
(\byear{2013})
\doiurl{10.1021/ja4072207}
\end{barticle}
\endbibitem

%%% 28
\bibitem[\protect\citeauthoryear{Poate and Saadatmand}{2002}]{IonImplantation}
\begin{barticle}
\bauthor{\bsnm{Poate}, \binits{J.M.}},
\bauthor{\bsnm{Saadatmand}, \binits{K.}}:
\batitle{{{Ion beam technologies in the semiconductor world (plenary)}}}.
\bjtitle{Rev. Sci. Instrum.}
\bvolume{73}(\bissue{2}),
\bfpage{868}--\blpage{872}
(\byear{2002})
\doiurl{10.1063/1.1428782}
\end{barticle}
\endbibitem

%%% 29
\bibitem[\protect\citeauthoryear{Kaner et~al.}{1987}]{CVD}
\begin{barticle}
\bauthor{\bsnm{Kaner}, \binits{R.B.}},
\bauthor{\bsnm{Kouvetakis}, \binits{J.}},
\bauthor{\bsnm{Warble}, \binits{C.E.}},
\bauthor{\bsnm{Sattler}, \binits{M.L.}},
\bauthor{\bsnm{Bartlett}, \binits{N.}}:
\batitle{{Boron-carbon-nitrogen materials of graphite-like structure}}.
\bjtitle{Mater. Res. Bull.}
\bvolume{22}(\bissue{3}),
\bfpage{399}--\blpage{404}
(\byear{1987})
\doiurl{10.1016/0025-5408(87)90058-4}
\end{barticle}
\endbibitem

%%% 30
\bibitem[\protect\citeauthoryear{Connell et~al.}{2015}]{Connell2015}
\begin{bchapter}
\bauthor{\bsnm{Connell}, \binits{S.H.}},
\bauthor{\bsnm{H{\"a}rtwig}, \binits{J.}},
\bauthor{\bsnm{Masvaure}, \binits{A.}},
\bauthor{\bsnm{Mavunda}, \binits{D.}},
\bauthor{\bsnm{{Tran Thi}}, \binits{T.N.}}:
\bctitle{{Towards a crystal undulator}}.
In: \beditor{\bsnm{Engelbrecht}, \binits{C.}},
\beditor{\bsnm{Karataglidis}, \binits{S.}} (eds.)
\bbtitle{{Proceedings of the 59th Annual Conference of the South African
  Institute of Physics (SAIP2014)}},
pp. \bfpage{169}--\blpage{174}.
\bpublisher{University of Johannesburg},
\blocation{Johannesburg}
(\byear{2015}).
\burl{https://events.saip.org.za/event/34/attachments/1143/1398/SAIP2014-169.pdf}
\end{bchapter}
\endbibitem

%%% 31
\bibitem[\protect\citeauthoryear{Brunet et~al.}{1998}]{BRUNET1998869}
\begin{barticle}
\bauthor{\bsnm{Brunet}, \binits{F.}},
\bauthor{\bsnm{Germi}, \binits{P.}},
\bauthor{\bsnm{Pernet}, \binits{M.}},
\bauthor{\bsnm{Deneuville}, \binits{A.}},
\bauthor{\bsnm{Gheeraert}, \binits{E.}},
\bauthor{\bsnm{Laugier}, \binits{F.}},
\bauthor{\bsnm{Burdin}, \binits{M.}},
\bauthor{\bsnm{Rolland}, \binits{G.}}:
\batitle{{The effect of boron doping on the lattice parameter of homoepitaxial
  diamond films}}.
\bjtitle{Diam. Relat. Mater.}
\bvolume{7}(\bissue{6}),
\bfpage{869}--\blpage{873}
(\byear{1998})
\doiurl{10.1016/S0925-9635(97)00316-6}
\end{barticle}
\endbibitem

%%% 32
\bibitem[\protect\citeauthoryear{Wojewoda et~al.}{2008}]{WOJEWODA20081302}
\begin{barticle}
\bauthor{\bsnm{Wojewoda}, \binits{T.}},
\bauthor{\bsnm{Achatz}, \binits{P.}},
\bauthor{\bsnm{Ortéga}, \binits{L.}},
\bauthor{\bsnm{Omnès}, \binits{F.}},
\bauthor{\bsnm{Marcenat}, \binits{C.}},
\bauthor{\bsnm{Bourgeois}, \binits{E.}},
\bauthor{\bsnm{Blase}, \binits{X.}},
\bauthor{\bsnm{Jomard}, \binits{F.}},
\bauthor{\bsnm{Bustarret}, \binits{E.}}:
\batitle{{Doping-induced anisotropic lattice strain in homoepitaxial heavily
  boron-doped diamond}}.
\bjtitle{Diam. Relat. Mater.}
\bvolume{17}(\bissue{7}),
\bfpage{1302}--\blpage{1306}
(\byear{2008})
\doiurl{10.1016/j.diamond.2008.01.040} .
\bcomment{Proceedings of Diamond 2007, the 18$^{\text{th}}$ European Conference
  on Diamond, Diamond-Like Materials, Carbon Nanotubes, Nitrides and Silicon
  Carbide}
\end{barticle}
\endbibitem

%%% 33
\bibitem[\protect\citeauthoryear{Sakai}{2011}]{MBE_SiGe}
\begin{bchapter}
\bauthor{\bsnm{Sakai}, \binits{A.}}:
\bctitle{{5 - Silicon–germanium (SiGe) Crystal Growth using Molecular Beam
  Epitaxy}}.
In: \beditor{\bsnm{Shiraki}, \binits{Y.}},
\beditor{\bsnm{Usami}, \binits{N.}} (eds.)
\bbtitle{Silicon–Germanium (SiGe) Nanostructures}.
\bsertitle{Woodhead Publishing Series in Electronic and Optical Materials},
pp. \bfpage{83}--\blpage{116}.
\bpublisher{Woodhead Publishing},
\blocation{Cham}
(\byear{2011}).
\doiurl{10.1533/9780857091420.2.83}
\end{bchapter}
\endbibitem

%%% 34
\bibitem[\protect\citeauthoryear{B{\"o}er and Pohl}{2020}]{DefectBook}
\begin{bbook}
\bauthor{\bsnm{B{\"o}er}, \binits{K.W.}},
\bauthor{\bsnm{Pohl}, \binits{U.W.}}:
\bbtitle{{Crystal Defects}},
pp. \bfpage{1}--\blpage{54}.
\bpublisher{Springer},
\blocation{Cham}
(\byear{2020}).
\doiurl{10.1007/978-3-319-06540-3_15-4}
\end{bbook}
\endbibitem

%%% 35
\bibitem[\protect\citeauthoryear{Tran~Thi et~al.}{2017}]{Diamond1}
\begin{barticle}
\bauthor{\bsnm{Tran~Thi}, \binits{T.N.}},
\bauthor{\bsnm{Morse}, \binits{J.}},
\bauthor{\bsnm{Caliste}, \binits{D.}},
\bauthor{\bsnm{Fernandez}, \binits{B.}},
\bauthor{\bsnm{Eon}, \binits{D.}},
\bauthor{\bsnm{H{\"{a}}rtwig}, \binits{J.}},
\bauthor{\bsnm{Barbay}, \binits{C.}},
\bauthor{\bsnm{Mer-Calfati}, \binits{C.}},
\bauthor{\bsnm{Tranchant}, \binits{N.}},
\bauthor{\bsnm{Arnault}, \binits{J.C.}},
\bauthor{\bsnm{Lafford}, \binits{T.A.}},
\bauthor{\bsnm{Baruchel}, \binits{J.}}:
\batitle{{{Synchrotron Bragg Diffraction Imaging Characterization of Synthetic
  Diamond Crystals for Optical and Electronic Power Device Applications}}}.
\bjtitle{J. Appl. Crystallography}
\bvolume{50}(\bissue{2}),
\bfpage{561}--\blpage{569}
(\byear{2017})
\doiurl{10.1107/S1600576717003831}
\end{barticle}
\endbibitem

%%% 36
\bibitem[\protect\citeauthoryear{Solov’yov et~al.}{2012}]{MBNExplorer}
\begin{barticle}
\bauthor{\bsnm{Solov’yov}, \binits{I.A.}},
\bauthor{\bsnm{Yakubovich}, \binits{A.V.}},
\bauthor{\bsnm{Nikolaev}, \binits{P.V.}},
\bauthor{\bsnm{Volkovets}, \binits{I.}},
\bauthor{\bsnm{Solov’yov}, \binits{A.V.}}:
\batitle{{MesoBioNano Explorer--a Universal Program for Multiscale Computer
  Simulations of Complex Molecular Structure and Dynamics.}}
\bjtitle{J. Comput. Chem.}
\bvolume{33}(\bissue{30}),
\bfpage{2412}--\blpage{2439}
(\byear{2012})
\doiurl{10.1002/jcc.23086}
\end{barticle}
\endbibitem

%%% 37
\bibitem[\protect\citeauthoryear{Sushko et~al.}{2019}]{MBNStudio}
\begin{barticle}
\bauthor{\bsnm{Sushko}, \binits{G.B.}},
\bauthor{\bsnm{Solov’yov}, \binits{I.A.}},
\bauthor{\bsnm{Solov’yov}, \binits{A.V.}}:
\batitle{{Modeling MesoBioNano Systems with MBN Studio Made Easy}}.
\bjtitle{J. Mol. Graph. Model.}
\bvolume{88},
\bfpage{247}--\blpage{260}
(\byear{2019})
\doiurl{10.1016/j.jmgm.2019.02.003}
\end{barticle}
\endbibitem

%%% 38
\bibitem[\protect\citeauthoryear{Gra{\v{z}}ulis
  et~al.}{2009}]{SiGeLatticeConstants}
\begin{barticle}
\bauthor{\bsnm{Gra{\v{z}}ulis}, \binits{S.}},
\bauthor{\bsnm{Chateigner}, \binits{D.}},
\bauthor{\bsnm{Downs}, \binits{R.T.}},
\bauthor{\bsnm{Yokochi}, \binits{A.F.T.}},
\bauthor{\bsnm{Quir{\'{o}}s}, \binits{M.}},
\bauthor{\bsnm{Lutterotti}, \binits{L.}},
\bauthor{\bsnm{Manakova}, \binits{E.}},
\bauthor{\bsnm{Butkus}, \binits{J.}},
\bauthor{\bsnm{Moeck}, \binits{P.}},
\bauthor{\bsnm{Le~Bail}, \binits{A.}}:
\batitle{{{Crystallography Open Database {--} an open-access collection of
  crystal structures}}}.
\bjtitle{J. Appl. Crystallogr.}
\bvolume{42}(\bissue{4}),
\bfpage{726}--\blpage{729}
(\byear{2009})
\doiurl{10.1107/S0021889809016690}
\end{barticle}
\endbibitem

%%% 39
\bibitem[\protect\citeauthoryear{Ethier and Lewis}{1992}]{SW_Ge-Ge}
\begin{barticle}
\bauthor{\bsnm{Ethier}, \binits{S.}},
\bauthor{\bsnm{Lewis}, \binits{L.J.}}:
\batitle{{Epitaxial Growth of Si1-xGex on Si(100)2 {\texttimes} 1: A
  Molecular-Dynamics Study}}.
\bjtitle{J. Mat. Res.}
\bvolume{7}(\bissue{10}),
\bfpage{2817}--\blpage{2827}
(\byear{1992})
\doiurl{10.1557/JMR.1992.2817}
\end{barticle}
\endbibitem

%%% 40
\bibitem[\protect\citeauthoryear{Stillinger and
  Weber}{1985}]{StillingerWeberOriginal}
\begin{barticle}
\bauthor{\bsnm{Stillinger}, \binits{F.H.}},
\bauthor{\bsnm{Weber}, \binits{T.A.}}:
\batitle{{Computer Simulation of Local Order in Condensed Phases of Silicon}}.
\bjtitle{Phys. Rev. B}
\bvolume{31}(\bissue{8}),
\bfpage{5262}--\blpage{5271}
(\byear{1985})
\doiurl{10.1103/PhysRevB.31.5262} .
\bcomment{{Erratum: Phys. Rev. B \textbf{33}(2), 1451 (1986)
  \href{https://doi.org/10.1103/PhysRevB.33.1451}{https://doi.org/10.1103/PhysRevB.33.1451}}}
\end{barticle}
\endbibitem

%%% 41
\bibitem[\protect\citeauthoryear{Dismukes et~al.}{1964}]{Dismukes1964}
\begin{barticle}
\bauthor{\bsnm{Dismukes}, \binits{J.P.}},
\bauthor{\bsnm{Ekstrom}, \binits{L.}},
\bauthor{\bsnm{Paff}, \binits{R.J.}}:
\batitle{{Lattice Parameter and Density in Germanium-Silicon Alloys}}.
\bjtitle{J. Phys. Chem.}
\bvolume{68}(\bissue{10}),
\bfpage{3021}--\blpage{3027}
(\byear{1964})
\doiurl{10.1021/j100792a049}
\end{barticle}
\endbibitem

%%% 42
\bibitem[\protect\citeauthoryear{Theodorou et~al.}{1994}]{Theodorou}
\begin{barticle}
\bauthor{\bsnm{Theodorou}, \binits{G.}},
\bauthor{\bsnm{Kelires}, \binits{P.C.}},
\bauthor{\bsnm{Tserbak}, \binits{C.}}:
\batitle{{Structural, electronic, and optical properties of strained
  ${\mathrm{Si}}_{1\mathrm{\ensuremath{-}}\mathit{x}}$${\mathrm{Ge}}_{\mathit{x}}$
  alloys}}.
\bjtitle{Phys. Rev. B}
\bvolume{50}(\bissue{24}),
\bfpage{18355}--\blpage{18359}
(\byear{1994})
\doiurl{10.1103/PhysRevB.50.18355}
\end{barticle}
\endbibitem

%%% 43
\bibitem[\protect\citeauthoryear{De~Salvador et~al.}{2000}]{De_Salvador}
\begin{barticle}
\bauthor{\bsnm{De~Salvador}, \binits{D.}},
\bauthor{\bsnm{Petrovich}, \binits{M.}},
\bauthor{\bsnm{Berti}, \binits{M.}},
\bauthor{\bsnm{Romanato}, \binits{F.}},
\bauthor{\bsnm{Napolitani}, \binits{E.}},
\bauthor{\bsnm{Drigo}, \binits{A.}},
\bauthor{\bsnm{Stangl}, \binits{J.}},
\bauthor{\bsnm{Zerlauth}, \binits{S.}},
\bauthor{\bsnm{M\"uhlberger}, \binits{M.}},
\bauthor{\bsnm{Sch\"affler}, \binits{F.}},
\bauthor{\bsnm{Bauer}, \binits{G.}},
\bauthor{\bsnm{Kelires}, \binits{P.C.}}:
\batitle{{Lattice parameter of
  ${\mathrm{Si}}_{1\ensuremath{-}x\ensuremath{-}y}{\mathrm{Ge}}_{x}{\mathrm{C}}_{y}$
  alloys}}.
\bjtitle{Phys. Rev. B}
\bvolume{61}(\bissue{19}),
\bfpage{13005}--\blpage{13013}
(\byear{2000})
\doiurl{10.1103/PhysRevB.61.13005}
\end{barticle}
\endbibitem

%%% 44
\bibitem[\protect\citeauthoryear{Vegard}{1921}]{Vegard1921}
\begin{barticle}
\bauthor{\bsnm{Vegard}, \binits{L.}}:
\batitle{{Die Konstitution der Mischkristalle und die Raumfüllung der Atome}}.
\bjtitle{Zeitschrift für Physik}
\bvolume{5}(\bissue{1}),
\bfpage{17}--\blpage{26}
(\byear{1921})
\doiurl{10.1007/BF01349680}
\end{barticle}
\endbibitem

%%% 45
\bibitem[\protect\citeauthoryear{Xu et~al.}{2017}]{SnGe_Discrepency_Exp}
\begin{barticle}
\bauthor{\bsnm{Xu}, \binits{C.}},
\bauthor{\bsnm{Senaratne}, \binits{C.L.}},
\bauthor{\bsnm{Culbertson}, \binits{R.J.}},
\bauthor{\bsnm{Kouvetakis}, \binits{J.}},
\bauthor{\bsnm{Menéndez}, \binits{J.}}:
\batitle{{Deviations from Vegard's law in semiconductor thin films measured
  with X-ray diffraction and Rutherford backscattering: The Ge1-ySny and
  Ge1-xSix cases}}.
\bjtitle{J. Appl. Phys.}
\bvolume{122}(\bissue{12}),
\bfpage{125702}
(\byear{2017})
\doiurl{10.1063/1.4996306}
\end{barticle}
\endbibitem

%%% 46
\bibitem[\protect\citeauthoryear{de~Gironcoli et~al.}{1991}]{Vegard_Disprep_1}
\begin{barticle}
\bauthor{\bsnm{Gironcoli}, \binits{S.}},
\bauthor{\bsnm{Giannozzi}, \binits{P.}},
\bauthor{\bsnm{Baroni}, \binits{S.}}:
\batitle{{Structure and thermodynamics of
  ${\mathrm{Si}}_{\mathit{x}}$${\mathrm{Ge}}_{1\mathrm{\ensuremath{-}}\mathit{x}}$
  alloys from ab initio Monte Carlo simulations}}.
\bjtitle{Phys. Rev. Lett.}
\bvolume{66}(\bissue{16}),
\bfpage{2116}--\blpage{2119}
(\byear{1991})
\doiurl{10.1103/PhysRevLett.66.2116}
\end{barticle}
\endbibitem

%%% 47
\bibitem[\protect\citeauthoryear{Tersoff}{1989}]{TersoffVegard}
\begin{barticle}
\bauthor{\bsnm{Tersoff}, \binits{J.}}:
\batitle{{Modeling solid-state chemistry: Interatomic potentials for
  multicomponent systems}}.
\bjtitle{Phys. Rev. B}
\bvolume{39}(\bissue{8}),
\bfpage{5566}--\blpage{5568}
(\byear{1989})
\doiurl{10.1103/PhysRevB.39.5566}
\end{barticle}
\endbibitem

%%% 48
\bibitem[\protect\citeauthoryear{Chizmeshya
  et~al.}{2003}]{SnGe_Discrepency_Exp_Theo}
\begin{barticle}
\bauthor{\bsnm{Chizmeshya}, \binits{A.V.G.}},
\bauthor{\bsnm{Bauer}, \binits{M.R.}},
\bauthor{\bsnm{Kouvetakis}, \binits{J.}}:
\batitle{{Experimental and Theoretical Study of Deviations from Vegard's Law in
  the SnxGe1-x System}}.
\bjtitle{Chem. Mert.}
\bvolume{15}(\bissue{13}),
\bfpage{2511}--\blpage{2519}
(\byear{2003})
\doiurl{10.1021/cm0300011}
\end{barticle}
\endbibitem

%%% 49
\bibitem[\protect\citeauthoryear{Jacob et~al.}{2007}]{Vegard_Approximation}
\begin{barticle}
\bauthor{\bsnm{Jacob}, \binits{K.T.}},
\bauthor{\bsnm{Raj}, \binits{S.}},
\bauthor{\bsnm{Rannesh}, \binits{L.}}:
\batitle{{Vegard's law: a fundamental relation or an approximation?}}
\bjtitle{Int. J. of Mater. Res.}
\bvolume{98}(\bissue{9}),
\bfpage{776}--\blpage{779}
(\byear{2007})
\doiurl{10.3139/146.101545}
\end{barticle}
\endbibitem

%%% 50
\bibitem[\protect\citeauthoryear{Adachi}{2009}]{Bowing_Parameter_book}
\begin{bbook}
\bauthor{\bsnm{Adachi}, \binits{S.}}:
\bbtitle{Properties of Semiconductor Alloys: Group-IV, III–V and II–VI
  Semiconductors}.
\bpublisher{Wiley},
\blocation{Chichester}
(\byear{2009}).
\doiurl{10.1002/9780470744383}
\end{bbook}
\endbibitem

%%% 51
\bibitem[\protect\citeauthoryear{Sushko et~al.}{2013}]{RelativisticMD}
\begin{barticle}
\bauthor{\bsnm{Sushko}, \binits{G.B.}},
\bauthor{\bsnm{Bezchastnov}, \binits{V.G.}},
\bauthor{\bsnm{Solov'yov}, \binits{I.A.}},
\bauthor{\bsnm{Korol}, \binits{A.V.}},
\bauthor{\bsnm{Greiner}, \binits{W.}},
\bauthor{\bsnm{Solov'yov}, \binits{A.V.}}:
\batitle{{Simulation of Ultra-relativistic Electrons and Positrons Channeling
  in Crystals with MBN Explorer}}.
\bjtitle{J. Comput. Phys.}
\bvolume{252},
\bfpage{404}--\blpage{418}
(\byear{2013})
\doiurl{10.1016/j.jcp.2013.06.028}
\end{barticle}
\endbibitem

%%% 52
\bibitem[\protect\citeauthoryear{Voigtländer}{2001}]{VOIGTLANDER2001127}
\begin{barticle}
\bauthor{\bsnm{Voigtländer}, \binits{B.}}:
\batitle{{Fundamental processes in Si/Si and Ge/Si epitaxy studied by scanning
  tunneling microscopy during growth}}.
\bjtitle{Surf. Sci. Rep.}
\bvolume{43}(\bissue{5}),
\bfpage{127}--\blpage{254}
(\byear{2001})
\doiurl{10.1016/S0167-5729(01)00012-7}
\end{barticle}
\endbibitem

%%% 53
\bibitem[\protect\citeauthoryear{Solov'yov et~al.}{2022}]{Stochastic_MBN}
\begin{barticle}
\bauthor{\bsnm{Solov'yov}, \binits{I.A.}},
\bauthor{\bsnm{Sushko}, \binits{G.}},
\bauthor{\bsnm{Friis}, \binits{I.}},
\bauthor{\bsnm{Solov'yov}, \binits{A.V.}}:
\batitle{{Multiscale modeling of stochastic dynamics processes with MBN
  Explorer}}.
\bjtitle{J. Comput. Chem.}
\bvolume{43}(\bissue{21}),
\bfpage{1442}--\blpage{1458}
(\byear{2022})
\doiurl{10.1002/JCC.26948}
\end{barticle}
\endbibitem

\end{thebibliography}

\end{document}